\documentclass[english]{article}
\usepackage[T1]{fontenc}
\usepackage[latin9]{inputenc}
\usepackage{geometry}
\usepackage{setspace}
\usepackage{microtype}
\makeatletter
\usepackage[margin=10pt,font=small,labelfont=bf,labelsep=endash]{caption}
\date{}

\makeatother

\usepackage{babel}
\begin{document}

\title{Introduction: Detectability of Future Earth}

\author{{\normalsize{}Jacob Haqq-Misra}\\
{\normalsize{}}\\
{\normalsize{}Blue Marble Space Institute of Science}\\
{\normalsize{}1001 4th Ave, Suite 3201, Seattle WA 98154}\\
{\normalsize{}jacob@bmsis.org}\\
{\normalsize{}}\\
{\normalsize{}Published in }\emph{\normalsize{}Futures}{\normalsize{}
(2019) 106: 1-3 as the introduction to the special issue on the }\\
{\normalsize{}``Detectability of Future Earth''}}

\maketitle
Changes in the Earth system since the advent of civilization include
alterations in continental land use, a rise in atmospheric greenhouse
gases, and exponential increases in energy consumption from industrialization
and modernization (IPCC 2013). The cumulative contribution of these
effects may delay or prevent the onset of the next glacial cycle (Berger
\& Loutre 2002), while trace elements of manufacturing may betray
the beginnings of industrialization in the geologic record (Waters
et al. 2014). The permanency of such changes has led some geologists
to propose that we are entering a new epoch known as the \textquotedblleft Anthropocene,\textquotedblright{}
which is characterized by geological and ecological developments that
result from the human influence (Steffen et al. 2007). As human civilization
ventures into the Anthropocene, it is becoming evident that our activities
are becoming increasingly intertwined with the evolution of the Earth
system.

The co-evolution of civilization and the planet will also affect the
observable characteristics of Earth from a distance. Astrobiologists
today are engaged in the search for spectroscopic \textquotedblleft biosignatures\textquotedblright{}
on extrasolar planets that would indicate the presence of surface
life, a feat that is becoming increasingly feasible with the next
generation of space telescopes (Schwieterman et al. 2018; Fujii et
al. 2018). Likewise, astronomers engaged in the search for extraterrestrial
intelligence (SETI) are seeking evidence of anomalous radio, optical,
infrared, or other electromagnetic signals that originate from extraterrestrial
technology (Wright 2017). Such efforts to detect extraterrestrial
life, and intelligence, are based upon the remotely detectable features
of life on Earth. Although scientists are prepared for a diverse assortment
of possible biosignatures and anti-biosignatures (Catling et al. 2018;
Meadows et al. 2018), the idea that life itself would exert a detectable
signature on a planet\textquoteright s atmosphere is based upon the
observation of such phenomenon on Earth. More exotic means of achieving
interstellar contact might exist in the distant future or far reaches
of the galaxy, but as far as we know today, the search for electromagnetic
evidence of life remains our best-available method for attempting
to discover extraterrestrial life. 

Understanding the evolution of Earth\textquoteright s detectability
across time, and into the future, provides important examples of biosignatures
and \textquotedblleft technosignatures\textquotedblright{} (Schneider
et al. 2010) that could plausibly exist elsewhere. Such scenarios
are also relevant as we think about the ethics of intentionally increasing
or decreasing Earth\textquoteright s overall detectability in order
to alter the probability of contact with extraterrestrial intelligence
(Vakoch 2011; Musso 2012). As human civilization is the only known
example of an energy-intensive civilization, our history and future
trajectories provide the basis for thinking about how to find life
elsewhere. This \emph{Futures} special issue features contributions
that consider the future evolution of the Earth system from an astrobiological
perspective, with the goal of exploring the extent to which anthropogenic
influence could be detectable across interstellar distances. 

Examining the detectability of future Earth requires that we first
consider the likely and possible trajectories for human civilization
as we enter the Anthropocene and contemplate becoming a spacefaring
civilization. In this special issue, Mullan and Haqq-Misra (2019)
examine trajectories of human population growth and energy use to
find that our technology may induce direct thermal heating (through
the second law of thermodynamics) by the year 2260. Even if population
growth stabilizes, these calculations indicate that exponential consumption
of energy could eventually lead our civilization to a point where
our energy use (fostered by the demand for higher standards of living)
would contribute to additional global warming beyond fossil fuel emissions.
The urgency of this situation is captured by David Grinspoon in his
book \emph{Earth in Human Hands}, which is reviewed in this issue
by Riggio (2019) for its challenge to think even more broadly about
the scale of change that is underway. Grinspoon\textquoteright s book
suggests that we are entering entirely uncharted territory in which
our co-evolution with Earth represents the planet awakening to self-awareness.
Finding solutions to some of these major global problems may come
with greater recognition of civilization itself as a planetary phenomenon. 

The second issue to consider about the detectability of future Earth
is whether any other technological civilizations have already navigated
through such trajectories. Forecasts are always prone to error and
speculation, but observational evidence of extraterrestrial civilizations
would provide an external basis of comparison for evaluating our options
for the future. Adam Frank discusses the probability of extraterrestrial
intelligence existing, as well our chances of finding any, in his
book \emph{Light of the Stars}, which is reviewed in this issue by
DeMarines (2019). Frank\textquoteright s book explores the relationship
between searching for technosignatures from energy-intensive civilizations
and projecting the possible fate of our own civilization. If it turns
out that energy-intensive civilizations are rare, then perhaps this
suggests that our own civilization cannot maintain its current trajectory
of consumption for much longer. Fundamental limits imposed by thermodynamics
and sustainability would restrict the rate of growth of any technological
civilization and slow its expansion into other planetary systems.
Mullan and Haqq-Misra (2019) likewise emphasize that slower-growth,
less energy-intensive civilizations are more likely to be prevalent
than any civilization that expands beyond the limits of its carrying
capacity. The lack of observations of any such energy-intensive civilizations
in the galaxy is a challenge for human civilization to adopt long-term
sustainable practices as we enter the Anthropocene.

The ongoing search for extraterrestrial life over about six decades
has found no tell-tale signs of any energy-intensive or sustainable-growth
civilizations. Such evidence could be forthcoming in coming years,
but this state of affairs may also indicate that extraterrestrial
intelligence does not exist at all. The argument famously known as
\textquotedblleft Fermi\textquoteright s paradox\textquotedblright{}
suggests that an advanced civilization could have rapidly spread across
the galaxy, so if they do exist we should have already seen them.
The absence of extraterrestrial visitors can be explained in many
ways, which DeVito (2019) explores mathematically in this special
issue. DeVito (2019) considers an expression for the rate of change
of communicative civilizations in the galaxy, constrained by the fact
that no such civilizations have yet been observed, and suggests that
one explanation to Fermi\textquoteright s paradox is that the number
of civilizations in the galaxy grows very slowly (logarithmically)
with time. Observationally addressing the Fermi paradox will require
continued efforts at SETI as well as the search for technosignatures,
which Simons and Haqq-Misra (2019) suggest in this issue would be
aided by a lunar observatory. The lack of an atmosphere on the moon
makes infrared wavelengths available for observations that would be
attenuated in terrestrial observatories by Earth\textquoteright s
atmosphere, while the presence of humans would make such an observatory
more continuous and permanent than an orbiting space telescope. Addressing
the question posed by Fermi\textquoteright s paradox---\textquotedblleft where
are they?\textquotedblright ---will help to constrain our expectations
for the lifetime of energy-intensive civilizations like our own.

Our civilization\textquoteright s population and energy growth may
exert a detectable influence on our climate, but our use of communicative
technology could be more easily limited through legislation, treaty,
or other agreement. As SETI surveys continues to seek evidence of
intentional transmissions or unintentional radio \textquotedblleft leakage\textquotedblright{}
from extraterrestrial civilizations, some critics have suggested that
our own transmissions emanating from Earth might constitute an existential
risk. In this issue, Haqq-Misra (2019) assesses the possible risk
that might accompany intentional transmissions into space and suggests
that our decision to become more radio-loud or radio-quiet in the
future depends upon our presumption of the benefits or harms that
would accompany extraterrestrial contact. In the absence of any such
decision to limit transmissions, our own radio leakage may otherwise
be detectable by advanced extraterrestrial astronomers observing in
our direction.

Underlying many of these concerns about Earth\textquoteright s detectability
is the longevity of our civilization. Thermodynamic constraints on
energy use as well as observational constraints from the Fermi paradox
indicate that our longevity depends upon our transition from an energy-intensive
to a sustainable civilization. Adopting policies of sustainable development
will require unprecedented effort on a global scale over many centuries.
As a way of inspiring this perspective, Som (2019) suggests in this
issue that efforts in early childhood psychology education could leverage
the \textquotedblleft overview effect\textquotedblright{} of Earth
as seen from space in order to promote cross-cultural exchange. Our
heterogeneous world often fails to recognize the limits of our global
commons, but finding a shared identity as Earthlings would be an important
step toward working toward a sustainable future and increasing the
longevity of our civilization.

Earth\textquoteright s future detectability depends intimately upon
the trajectory of our civilization over the coming centuries. This
collection of papers emphasizes the connection between the unfolding
future of the Anthropocene with the search for extraterrestrial civilizations.
Our rate of energy consumption will characterize the extent to which
our energy-intensive society exerts direct influence on climate, which
in turn may limit the ultimate lifetime of our civilization. If the
answer to Fermi\textquoteright s question is that we are alone, so
that our civilization represents the only form of intelligent life
in the galaxy (or even the universe), then our responsibility to survive
is even greater. If we do find evidence of another civilization on
a distant exoplanet, then at least we will know that our trajectory
can be managed. But as long as our searches turn up empty, we must
stay vigilant to keep our future secure. 

\subsubsection*{References}

{\footnotesize{}Berger, A., \& Loutre, M. F. (2002). An exceptionally
long interglacial ahead? }\emph{\footnotesize{}Science}{\footnotesize{},
297(5585), 1287-1288.}\\
{\footnotesize{}}\\
{\footnotesize{}Catling, D. C., Krissansen-Totton, J., Kiang, N. Y.,
Crisp, D., et al. (2018). Exoplanet biosignatures: A framework for
their assessment. }\emph{\footnotesize{}Astrobiology}{\footnotesize{},
18(6), 709-738.}\\
{\footnotesize{}}\\
{\footnotesize{}DeMarines, J. (2019). Light of the Stars: Alien Worlds
and the Fate of the Earth, by Adam Frank. }\emph{\footnotesize{}Futures}{\footnotesize{},
106: 20.}\\
{\footnotesize{}}\\
{\footnotesize{}DeVito, C. (2019). On the meaning of Fermi's paradox.
}\emph{\footnotesize{}Futures}{\footnotesize{}, 106: 21-23..}\\
{\footnotesize{}}\\
{\footnotesize{}Fujii, Y., Angerhausen, D., Deitrick, R., Domagal-Goldman,
S., et al. (2018). Exoplanet biosignatures: Observational prospects.
}\emph{\footnotesize{}Astrobiology}{\footnotesize{}, 18(6), 739-778.}\\
{\footnotesize{}}\\
{\footnotesize{}Meadows, V. S., Reinhard, C. T., Arney, G. N., Parenteau,
M. N., et al. (2018). Exoplanet biosignatures: Understanding oxygen
as a biosignature in the context of its environment. }\emph{\footnotesize{}Astrobiology}{\footnotesize{},
18(6), 630-662.}\\
{\footnotesize{}}\\
{\footnotesize{}Mullan, B. \& Haqq-Misra, J. (2019). Population growth,
energy use, and the implications for the search for extraterrestrial
intelligence. }\emph{\footnotesize{}Futures}{\footnotesize{}, 106:
4-17..}\\
{\footnotesize{}}\\
{\footnotesize{}Musso, P. (2012). The problem of active SETI: An overview.
}\emph{\footnotesize{}Acta Astronautica}{\footnotesize{}, 78, 43-54.}\\
{\footnotesize{}}\\
{\footnotesize{}Haqq-Misra, J. (2019). Policy options for the radio
detectability of Earth. }\emph{\footnotesize{}Futures}{\footnotesize{},
106: 33-36.}\\
{\footnotesize{}}\\
{\footnotesize{}IPCC, 2013: Summary for policymakers. In: }\emph{\footnotesize{}Climate
Change 2013: The Physical Science Basis. Contribution of Working Group
I to the Fifth Assessment Report of the Intergovernmental Panel on
Climate Change}{\footnotesize{} {[}Stocker, T.F., D. Qin, G.-K. Plattner,
M. Tignor, S.K. Allen, J. Boschung, A. Nauels, Y. Xia, V. Bex and
P.M. Midgley (eds.){]}. Cambridge University Press, Cambridge, United
Kingdom and New York, NY, USA}\\
{\footnotesize{}}\\
{\footnotesize{}Riggio, G. (2019). Earth in Human Hands, by David
Grinspoon. }\emph{\footnotesize{}Futures}{\footnotesize{}, 106: 18-19.}\\
{\footnotesize{}}\\
{\footnotesize{}Schneider, J., Léger, A., Fridlund, M., White, G.
J., et al. (2010). The far future of exoplanet direct characterization.
}\emph{\footnotesize{}Astrobiology}{\footnotesize{}, 10(1), 121-126.}\\
{\footnotesize{}}\\
{\footnotesize{}Schwieterman, E. W., Kiang, N. Y., Parenteau, M. N.,
Harman, C. E., et al. (2018). Exoplanet biosignatures: A review of
remotely detectable signs of life. }\emph{\footnotesize{}Astrobiology}{\footnotesize{},
18(6), 663-708.}\\
{\footnotesize{}}\\
{\footnotesize{}Steffen, W., Crutzen, P. J., \& McNeill, J. R. (2007).
The Anthropocene: Are humans now overwhelming the great forces of
nature. }\emph{\footnotesize{}AMBIO: A Journal of the Human Environment}{\footnotesize{},
36(8), 614-621.}\\
{\footnotesize{}}\\
{\footnotesize{}Simons, S. \& Haqq-Misra, J. (2019). A trip to the
moon might constrain the Fermi paradox. }\emph{\footnotesize{}Futures}{\footnotesize{},
106: 24-32.}\\
{\footnotesize{}}\\
{\footnotesize{}Som, S. (2019). Common identity as a step to civilization
longevity. }\emph{\footnotesize{}Futures}{\footnotesize{}, 106: 24-32.}\\
{\footnotesize{}}\\
{\footnotesize{}Vakoch, D. A. (2011). Asymmetry in Active SETI: A
case for transmissions from Earth. }\emph{\footnotesize{}Acta Astronautica}{\footnotesize{},
68(3-4), 476-488.}\\
{\footnotesize{}}\\
{\footnotesize{}Waters, C. N., Zalasiewicz, J. A., Williams, M., Ellis,
M. A., \& Snelling, A. M. (2014). A stratigraphical basis for the
Anthropocene? }\emph{\footnotesize{}Geological Society, London, Special
Publications}{\footnotesize{}, 395, SP395-18.}\\
{\footnotesize{}}\\
{\footnotesize{}Wright J. (2017) Exoplanets and SETI. In: Deeg H.,
Belmonte J. (eds) }\emph{\footnotesize{}Handbook of Exoplanets}{\footnotesize{}.
Springer, Cham.}{\footnotesize\par}
\end{document}